\documentclass[%
 reprint,
 amsmath,amssymb,
 aps,
]{revtex4-1}

\usepackage{graphicx}
\usepackage{dcolumn}
\usepackage{bm}

\usepackage{color}
\usepackage[version=3]{mhchem}

\begin{document}

\preprint{APS/123-QED}

\title{The dynamical and thermodynamical origin of dissipative chaos}


\author{Feng Zhang$^1$}
\affiliation{$^1$State Key Laboratory of Electroanalytical Chemistry, Changchun Institute of Applied Chemistry, Chinese Academy of Sciences, Changchun, Jilin 130022, China}

\author{Liufang Xu$^2$}
\affiliation{$^2$College of Physics, Jilin University, Changchun 130012, China}

\author{Jin Wang$^3$}
\email{jin.wang.1@stonybrook.edu}
\affiliation{$^3$Department of Chemistry and of Physics and Astronomy, State University of New York at Stony Brook, Stony Brook, NY 11794-3400, USA}

\date{\today}


\begin{abstract}
Chaos is usually referred to the sensitivity to initial conditions in which the nonlinearity plays a crucial role.
Beyond such a mathematical description, the understanding of the underlying physical origin of the chaos is still not very clear.
Here we study the dissipative chaos from the perspective of the nonequilibrium dynamics.
This was not fully investigated in the traditional chaos theory, despite of the Lorenz's original discovery of chaos from the nonequilibrium atmosphere.
We found that the nonequilibriumness as the degree of detailed balance breaking can be quantified by the appearance of the steady state probability flux in the state space.
We uncovered that the dynamical origin of the onset and offset of the dissipative chaos such as Lorentz attractor is from the sudden appearance and disappearance of such nonequilibrium flux.
We also uncovered that the dissipation associated with the flux quantified by the entropy production rate gives the thermodynamical origin of dissipative chaos.
The sharp changes in the degree of nonequilibriumness by the flux and the entropy production rate also provide alternative quantitative indicators for the onset and offset of the dissipative chaos.
\end{abstract}

\maketitle


\section{Introduction}

Chaos is usually refereed to the behavior of the dynamical systems highly sensitivity to the initial conditions\cite{Ott}.
The butterfly effect is a popular metaphor for the chaos.
A tiny difference in the initial setup of a nonlinear system can lead to totally different evolutionary outcome.
Chaos can often be found in both physical and biological world\cite{Gleick}.
The atmosphere dynamics\cite{Lorenz1963} and the population growth in ecology\cite{May1976} are some typical examples.
The chaos seemingly originate from the noises or the environmental fluctuations.
However, researches have found that chaos also emerge in the deterministic systems without the fluctuations\cite{Lorenz1963,May1976}, and simple rules can create complex behavior\cite{May1976}.
Researchers have also found various markers for capturing this crucial feature of chaos\cite{Ott}, such as the Lyapunov exponents.
Since Lorenz's discovery from the atmosphere convection\cite{Lorenz1963}, chaos has been studied intensively for almost sixty years.

The early explorations of the chaos started from the Hamiltonian system without dissipation in particular for the celestial mechanics, for instance the gravitational many-body systems and Henon-Heiles system\cite{Henon1964}, which are often termed as Hamiltonian chaos.
The stability of the solar system is a fundamental issue for the Hamiltonian chaos. The subsequent development involved with the KAM theorem which provides an explanation for the stability with respect to the weak perturbations\cite{Ott}.
The trajectories of an integrable Hamiltonian system are constrained by KAM surfaces. The destruction of the surfaces drives the system to chaos with the increase of perturbations.
However, a dynamical system is more often dissipative out of equilibrium. We can term the chaos in dissipation systems as dissipative chaos.

A phenomenological distinction between dissipative and Hamiltonian chaos comes from the response of the system to the perturbation.
The phases of the orbits in a Hamiltonian system can be destroyed by the perturbation, while the dissipative system can sustain its distinct nature by contracting the dynamics to a bounded set called strange attractors originated from the nonequilibrium conditions.
A biological organism can benefit from the attractor to acquire the stable physiological functions, for instance the activities of the heart\cite{Poon1997} and the brains\cite{Rabinovich1998,Barak2013}.
The dissipative chaos is ubiquitous compared to the Hamiltonian chaos, and possesses intriguing consequences associated with the nonequilibriumness.

The Lorenz's discovery of chaos comes from the nonequilibrium dissipative system, a temperature difference imposed atmosphere\cite{Lorenz1963}.
It is interesting to observe that the atmosphere evolves into chaotic stage as the temperature difference increases.
The further evolution out of control gives rise to the turbulence in the atmosphere. The exploration of the dissipative chaos was also motivated from the researches on the turbulence in the fluid\cite{Ruelle1971,Newhouse1978}.
%
%
%
The crucial feature of chaos, i.e., the sensitivity to initial conditions, has also promoted the development of the cryptography\cite{Matthews1989}.
On the other hand, certain control methods have also been suggested to eliminate the unpredictable risks induced by the chaos\cite{Yorke1990,Pyragas1992}.

Different scenarios have  been suggested for the generation of chaos\cite{Eckmann1981}.
The sequence of bifurcations as a route to chaos\cite{Schuster} has been proposed in several fields, such as hydrodynamics\cite{Libchaber1982} and electronics\cite{Linsay1981}.
Amazingly, seemingly uncorrelated fields share the same constants, called Feigenbaum constants, from a mathematical model of animal populations\cite{Feigenbaum1978}.
An universality thus appears in the route to chaos and is independent of the special model\cite{Gleick}.

The nonlinearity is conventionally thought as the origin of the chaos\cite{Schuster}.
If the dynamics is known, we can investigate how a system transforms into chaotic stage, through the nonlinear actions in certain parameter regimes.
However, it is difficult to know the specific dynamics to investigate the chaos except for some simple systems.
This is especially relevant for the large spatiotemporal scales system involving large number of interacting elements.
In addition, every system possesses its own nonlinear features for creating the chaos.
Accordingly, the generations of chaos are realized in different scenarios\cite{Eckmann1981} for different systems without an uniform mechanism.
From the Feigenbaum's studies, one can ask if there is an universal mechanism for the chaos beyond just the nonlinearity?

An alternative approach in understanding the chaos comes from the thermodynamical perspective.
A complex system is subjected inevitably to the thermodynamical laws, especial under the large spatiotemporal scale systems.
The macroscopic evolution will eventually come to the equilibrium.
For instance, the convection vanishes in the atmosphere without the temperature difference.
From the thermodynamical perspective, the nonequilibrium condition should be responsible for the universal origin of the chaos.
In fact, the nonequilibriumness is implicitly coupled with the parameters regulating the specific nonlinear effects.
Looking back the Lorenz's discovery of chaos, the Rayleigh number plays an important role for buoyancy-driven flow\cite{Lorenz1963}.
The Rayleigh number can be considered as a measure of the nonequilibriumness, i.e., the temperature difference between the bottom and the top of atmosphere.
However, despite of the Lorenzs original discovery of chaos from the nonequilibrium atmosphere, it is still a grand challenge to understand the universal origin of chaos as well as the associated underlying physical mechanism, even after the intensive studies of chaos for nearly sixty years.
This is the motivation of our current work to explore the physical origin of the dissipative chaos.

In this study, we uncovered the universal origin of the onset and offset of dissipative chaos.
We found that the nonequilibriumness measuring the degree of detailed balance breaking can be quantified by the appearance of steady state probability flux in the state space.
We uncovered that the dynamical origin of the onset and offset of the dissipative chaos is from the sudden appearance and disappearance of such nonequilibrium flux.
We also uncovered that the thermodynamic dissipation associated with the flux quantified by the entropy production rate gives the thermodynamic origin of dissipative chaos.
The sharp changes in the degree of nonequilibriumness by the flux and the entropy production rate also provide alternative quantitative indicators for the onset and offset of dissipative chaos.
Our study provides not only an understanding on the physical origin of dissipative chaos, but also an insight on the origin of the nonequilibrium phase transitions.

\section{The driving force for the chemical reactions}

To study the dissipative chaos, we will explore a chemical Lorentz system\cite{Samardzija1989} as an example mimicking the behavior of classical Lorenz model\cite{Lorenz1963}. Using the chemical Lorentz system for study has an advantage that the physical driving force for the dynamics can be identified clearly as we will see later.
The atmosphere evolves into chaotic stage as the temperature difference increases.
The similar effect can also be found in the chemical systems\cite{Samardzija1989}.
The chemical systems often possess inherently the nonlinear features from the law of mass action which can give rise to the chaotic dynamics.

Imagine that there is a chemical reaction system embedded between some particle reservoirs.
The chemical system consists of $M$ species and $N$ reactions and the reactions are described by the general form
\begin{eqnarray}
\sum\nolimits_m v^+_{nm} X_m   \ce{<=>[k^+_m][k^-_m]}   \sum\nolimits_m v^-_{nm} X_m .
\end{eqnarray}
In coarse-grained description, the state of the system is determined by the population concentrations $x_m$.
We can introduce the progress variable $\xi_n$ for counting the reaction process for $n$-th reaction. The progress variable increases $\xi_n \rightarrow \xi_n+1$ with the one step $n$-th forward reaction.
The inharmony among different pathways results in the accumulation in the node species with the rate
\begin{eqnarray}\label{kinematic}
\dot{x}_m = \sum\nolimits_m v_{nm} \dot{\xi}_n ,
\end{eqnarray}
where the stoichiometric coefficients $v_{nm} = \partial x_m / \partial \xi_n = v^+_{nm} - v^-_{nm}$ reflect the stoichiometric structure of the reaction networks.
The stoichiometric coefficient $v_{nm} < 0$ for reactants and $v_{nm} > 0$ for products.
The nonlinearity is introduced through the change rate of progress variable $\dot{\xi}_n$ determined by the law of mass action.

From the thermodynamical perspective, a particle in a chemical system is driven by chemical potential $\mu$.
The chemical potential can be viewed as an effective statistical mechanical pressure on the particle creating an effective force which is called affinity defined as
\begin{eqnarray}\label{dynamic1}
A_n = - \sum\nolimits_m v_{nm} \mu_m .
\end{eqnarray}
%
%
A non-zero affinity breaks the reaction equilibrium of $n$-th reaction 
\begin{eqnarray}\label{dynamic2}
A_n = \ln(J_n^+/J_n^-)
\end{eqnarray}
and creates the reaction flow 
\begin{eqnarray}\label{dynamic3}
\dot{\xi}_n = J_n^+ - J_n^- ,
\end{eqnarray}
where $J_n^+$ and $J_n^-$ are the forward and backward reaction fluxes respectively.
Note the signs of stoichiometric coefficients, i.e., $v_{nm} < 0$ for reactants and $v_{nm} > 0$ for products.
So, the affinities $A_n$ reflect in fact the chemical potential difference $\delta_n \mu$ between the reactant and product of $n$-th reaction.

It is worth to note Eq.\ref{kinematic} addresses how the reactions proceed, while Eq.\ref{dynamic1} associated with Eq.\ref{dynamic2} and Eq.\ref{dynamic3} shows why the reactions proceed.
The causal relationship can be written as the implicit expression
\begin{eqnarray}\label{implicit}
\dot{x}_m = \sum\nolimits_m v_{nm} \dot{\xi}_n[A_n(\delta_n \mu)] .
\end{eqnarray}
In this way, the temperature difference imposed on the atmosphere can now be replaced by the chemical potential difference in chemical systems.
%
%
The latter can be viewed as an effective pressure on the particles generating the chemical reaction flows analogous to the electric voltage driving the current in an electric circuit\cite{Schnakenberg1976}.

The reactant and product can be in different environments with different chemical potentials to sustain the nonequilibriumness.
We can introduce an overall chemical potential difference between the reactant and the product environment reservoirs in the form
\begin{eqnarray}
\Delta\mu = \sum\nolimits_i \mu_{i} - \sum\nolimits_j \alpha_j\mu_{j}
\end{eqnarray}
where $\mu_{i}$ stands for the chemical potential of the reactant reservoirs and  $\mu_{j}$ stands for the product reservoirs in the similar way.
The coefficients $\alpha_j$ reflect the reaction structure.
The chemical potential can be written in the form $\mu_k = \mu^0_k+\ln x_k$ with concentration $x_k$, where $\mu^0_k$ is the chemical potential of pure species $k$.
The chemical potential difference will possess complex form due to different features of species.

For a forward dominated reaction system, we can simplify the chemical potential difference into a relatively compact structure.
In such a system, the backward reactions occur under very small rates, so that the equilibriumness can only be achieved with very small amount of reactants.
In other words, the chemical equilibrium $\Delta\mu \rightarrow 0$ is sustained by very small concentration of reactants.
In contrast, the finite concentrations of reactants contribute to additional chemical potential difference
\begin{eqnarray}
\Delta\mu = \sum\nolimits_i \ln r_i
\end{eqnarray}
where $r_i$ is the concentration of the $i$-th reactant reservoir.
It breaks the chemical equilibrium to create chemical reaction flows, similar to a higher temperature is imposed on the bottom of the atmosphere to create the buoyancy.
The chemical potential difference increases with the density of reactants, while the products do not contribute to such difference. 
%

\section{Model description}

\begin{figure}[ht]
\centering\includegraphics[width=2in]{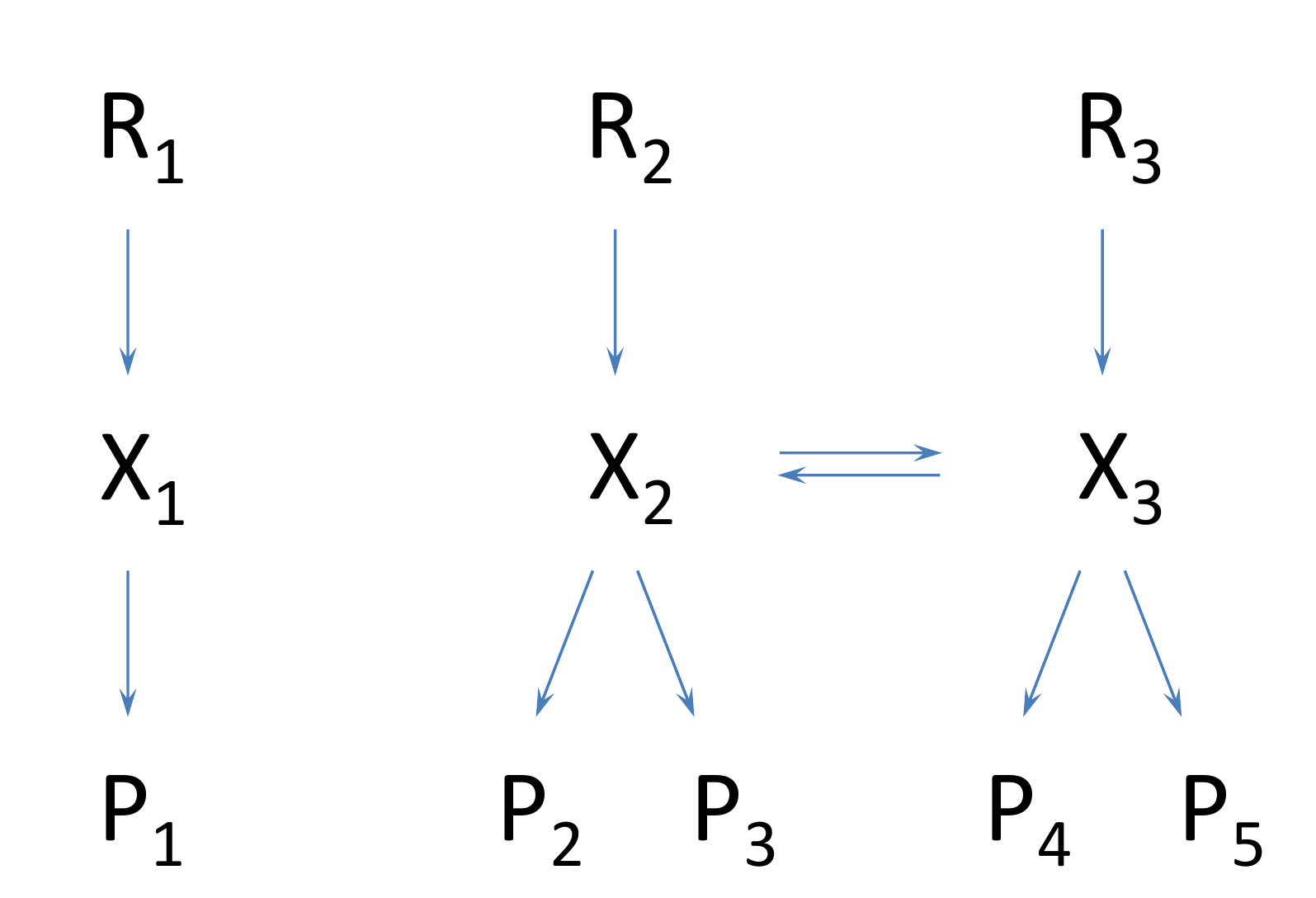}
\caption{
Sketch for the pathways of species transitions.
}\label{Sketch}
\end{figure}

Let us consider a chaotic chemical system\cite{Samardzija1989} mimicking the classical Lorenz model\cite{Lorenz1963} described by a series of reactions(details see Appendix A)
\begin{eqnarray}
\ce{ R1  +  X1  +  X2         ->[k_1]         2X1  +  X2             } ,     \\ \nonumber
\ce{ R2  +  X1  +  X2         ->[k_2]          X1  + 2X2             } ,     \\ \nonumber
\ce{ R3         +  X3         ->[k_3]                      2X3       } ,     \\ \nonumber
\ce{        X1  +  X2  +  X3  ->[k_4]          X1        + 2X3       } ,     \\ \nonumber
\ce{               X2  +  X3  ->[k_5]                2X2             } ,     \\ \nonumber
\ce{       2X1                ->[k_6]                             P1 } ,     \\ \nonumber
\ce{              2X2         ->[k_7]                             P2 } ,     \\ \nonumber
\ce{               X2         ->[k_8]                             P3 } ,     \\ \nonumber
\ce{        X1         +  X3  ->[k_9]          X1               + P4 } ,     \\ \nonumber
\ce{                     2X3  ->[k_{10}]                          P5 } .
\end{eqnarray}
%
%
%
The pathways for the chemical species transitions relevant to such reactions are sketched as in Fig.\ref{Sketch}.
Such a chemical system can be formulated as the system $\mathbf{X}$ immersed between a particle reservoirs $\mathbf{R}$ with higher chemical potentials and particle reservoirs $\mathbf{P}$ with lower chemical potentials.
The chemical potential difference in this case is given as
\begin{eqnarray}
\Delta\mu = \ln r_1
\end{eqnarray}
where $r_1$ is the concentration of the first reactant reservoirs.

\begin{figure*}[ht]
\centering\includegraphics[width=4in]{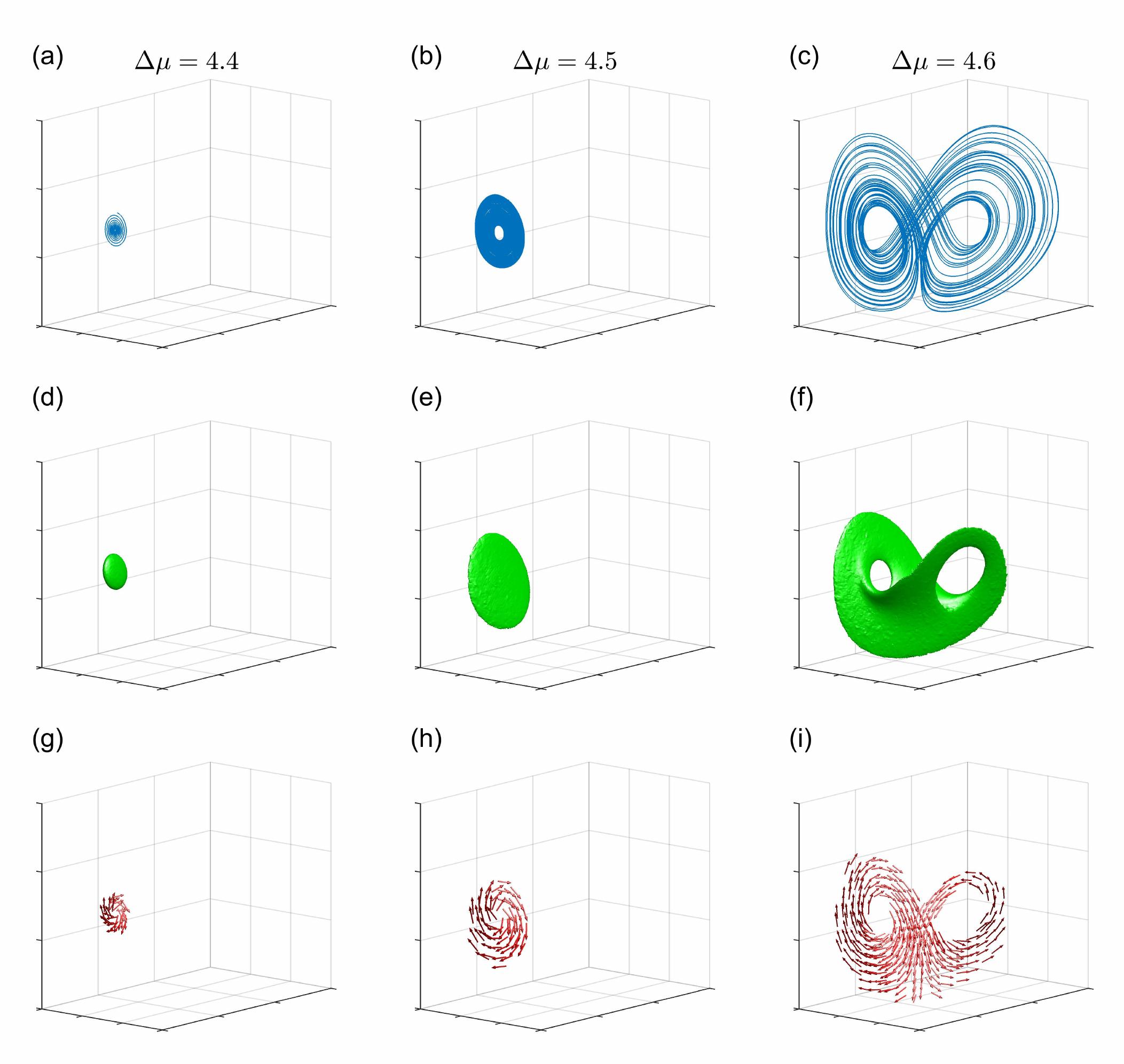}
\caption{
Different stages with the chemical potential difference. All the subgraphs share the same size of the coordinate axis frame, where the calibrations are removed for the concision.
(a)-(c), The deterministic trajectories. (d)-(f), The isosurface for the intrinsic potential(details see Appendix C). (g)-(i), The normalized intrinsic flux.
The converged trajectory exhibits a mono-stable phase in a small difference in the chemical potential as in (a). The system state is attracted into a narrow basin by the negative gradient of potential as in (d), and the attracted orbit is curved by the influence of flux as in (g).
The increased chemical potential difference pushes the system to enter into a critical point of phase transition. The system state can not reach the equilibrium state in finite time due to the critical slowing down, so that there is an small cavity in the finite-time trajectory as in (b). In this case, the system possesses flat basin of attraction corresponding to a large isosurface as in (e). The flux does not undergo a large transition as in (h) compared with (g).
The system enters into a chaotic phase with further increased chemical potential difference as in (c). The potential landscape loses its own gradient effect $-\nabla\phi$ in the basin of attractor as in (f). After the system fell into the basin of attraction, the system dynamics is governed by the intrinsic flux $\mathbf{V}$ as in (i). The system enters into an endless chaotic evolution driving by the intrinsic flux.
}\label{Fig01}
\end{figure*}

\section{Master Equation for Stochastic Dynamics and Potential-Flux Representation}

In a coarse-grained description, the system state is determined by the total particle numbers $\Omega\mathbf{x}$ of each species where $\Omega$ is the volume of container.
From microscopic perspective, these chemical reactions are actually stochastic processes.
The $n$-th reaction occurs with the probability $W_n(\Omega\mathbf{x},\Omega\mathbf{x}-\mathbf{v}_n)$ from state $\Omega\mathbf{x}-\mathbf{v}_n$ to $\Omega\mathbf{x}$ per unit time.
Here, the stoichiometric coefficients are written as the vector form $\mathbf{v}_n$, where the components $v_{ni}$ is the stoichiometric coefficient of $i^{th}$ species in $n^{th}$ reaction.
Due to the intrinsic statistical number fluctuations, it is necessary to introduce a statistical description.
Such a stochastic system can be described by the master equation
\begin{eqnarray}
\dot{P}(\mathbf{X})
=
\sum\nolimits_n [W_n(\mathbf{X},\mathbf{X}-\mathbf{v}_n) P(\mathbf{X}-\mathbf{v}_n)   \\ \nonumber
-  W_n(\mathbf{X}+\mathbf{v}_n,\mathbf{X}) P(\mathbf{X})]
\end{eqnarray}
where $P(\mathbf{X})$ is the probability of the system staying in the state $\mathbf{X}$.

The concentration $\mathbf{x}$ is usually an invariant with respect to the volume of container $\mathbf{\Omega}$. For instance, in the thermodynamic limit or macroscopic limit, the concentrations hold fixed $\mathbf{X}/\Omega = \mathrm{constant}$, while $\mathbf{X}\rightarrow\infty$ and $\Omega\rightarrow\infty$.
So, it is easy to obtain a scaling transform for the master equation\cite{vanKampen}
\begin{eqnarray}
\dot{\rho}(\mathbf{x})
=
\Omega \sum\nolimits_n [ w_n\left(\mathbf{x},\mathbf{x}-\frac{\mathbf{v}_n}{\Omega}\right) \rho\left(\mathbf{x}-\frac{\mathbf{v}_n}{\Omega}\right)   \\ \nonumber
-  w_n\left(\mathbf{x}+\frac{\mathbf{v}_n}{\Omega},\mathbf{x}\right) \rho\left(\mathbf{x}\right) ]
\end{eqnarray}
with the scaling maps
\begin{eqnarray}
\mathbf{X}       &=&  \Omega \mathbf{x} ,     \\ \nonumber
P(\mathbf{X})    &=&  \Omega^{-1} \rho(\mathbf{x}) ,     \\ \nonumber
W_n(\mathbf{X})  &=&  \Omega w_n(\mathbf{x}).
\end{eqnarray}
A large system $\Omega\gg1$ can be described by the Fokker-Planck equation\cite{Gardiner,Risken}
\begin{eqnarray}
\dot{\rho}(\mathbf{x})  =  - \nabla\cdot\mathbf{J}
\end{eqnarray}
with the probability flux
\begin{eqnarray}\label{flux}
\mathbf{J} = \mathbf{F} \rho - \Omega^{-1} \nabla\cdot(\mathbf{D}\rho) .
\end{eqnarray}
At the long times, the probability distribution is at a steady state $\rho_\mathrm{ss}(\mathbf{x})$ along with a divergence-free steady state probability flux $\mathbf{J}_\mathrm{ss}$ possessing a rotational nature
\begin{eqnarray}
\nabla\cdot\mathbf{J}_\mathrm{ss} = 0
\end{eqnarray}
where the nonzero $\mathbf{J}_\mathrm{ss}$ represents the net input/output and measures the degree of the detailed balance breaking.
The drift term $\mathbf{F}$ represents the driving forces for the deterministic species dynamics
\begin{eqnarray}\label{F}
\dot{x}_i = F_i(\mathbf{x}) = \sum\nolimits_n v_{ni} w_n(\mathbf{x})
\end{eqnarray}
and the diffusion term $\mathbf{D}$ represents the correlation of fluctuations
\begin{eqnarray}\label{D}
D_{ij}(\mathbf{x}) = \sum\nolimits_n v_{ni}v_{nj} w_n(\mathbf{x}) / 2 ,
\end{eqnarray}
where the transition rate densities
\begin{eqnarray}
w_n(\mathbf{x}) = \Omega^{-1}W_n(\Omega\mathbf{x})
\end{eqnarray}
are usually expressed by the law of mass action.

It may seem confusing why we want to go to the probabilistic description rather than the trajectory description.
In fact, the probability is a global measure rather than the local measure of the dynamics since it gives rise to the weight of all the states.
In addition, the trajectory follows nonlinear equation and is unpredictable, while the probability evolution follows a linear dynamics and is therefore predictable.
Therefore, the probabilistic description provides an alternative approach to understand the origin of chaos as described further in the following discussions.

The physical meaning is clear by rewriting the flux expression(Eq.\ref{flux}) as the form
\begin{eqnarray}
\mathbf{F}  =  \Omega^{-1} \mathbf{D} \cdot \nabla\ln\rho_\mathrm{ss}  +  \mathbf{J}_\mathrm{ss} / \rho_\mathrm{ss}  -  \nabla\cdot(\Omega^{-1}\mathbf{D}).
\end{eqnarray}
Note that the last term vanishes in a large system limit, $\Omega \gg 1$.
The driving force in the dynamics is from both the gradient of the steady state probability landscape and a rotational force provided by a curl steady state probability flux. For equilibrium system, detailed balance is reflected by the zero value of the flux. The dynamics is determined by the gradient of the equilibrium landscape. In contrast, for nonequilibrium system, the dynamics is determined by both the gradient of the probability landscape and the flux.

Specifically, the deterministic species dynamics in the limit of zero fluctuations can be written in an autonomous form from the law of mass action(details see Appendix A)
\begin{eqnarray}
\dot{x}_1
&=&
k_1 \exp(\Delta\mu) x_1 x_2	- 2k_6 x_1^2,                          \\ \nonumber
\dot{x}_2
&=&
k_2 r_2x_1x_2 - k_4x_1x_2x_3 + k_5x_2x_3	- 2k_7x_2^2	-k_8x_2	,		\\ \nonumber
\dot{x}_3
&=&
k_3 r_3x_3 + k_4x_1x_2x_3 - k_5x_2x_3 - k_9x_1x_3 - 2k_{10}x_3^2 .
\end{eqnarray}
%
%
%
%
The main behaviors of the system dynamics can be captured by the deterministic trajectories as in Fig.\ref{Fig01}(a)-(c) at different stages.
The increase in the chemical potential difference triggers an onset transition to chaos.
As the chemical potential difference increases, a mono-stable system (Fig.\ref{Fig01}(a)) enters into a chaotic phase(Fig.\ref{Fig01}(c)) through a transition stage(Fig.\ref{Fig01}(b)).
The dynamical trajectories converge into a fixed point in a mono-stable phase(Fig.\ref{Fig01}(a)).
In contrast, the trajectories wind  around between two branches in the chaotic phase(Fig.\ref{Fig01}(c)).
The initially infinitesimally closed trajectories separate exponentially.
The system behavior is highly sensitive to the initial conditions.

\section{Thermodynamical limit}

In the thermodynamical limit $\Omega\rightarrow\infty$, the steady-state solution $\rho_\mathrm{ss}$ of the Fokker-Planck equation can be written in a form of Wentzel-Kramers-Brillouin(WKB) expansion\cite{Graham1984}
\begin{eqnarray}
\rho_\mathrm{ss}  =  \exp\left( \Omega \phi + \sum\nolimits_{k=1}^\infty \Omega^{1-k} \phi_k \right) .
\end{eqnarray}
By inserting the expansion into the steady-state Fokker-Planck equation
\begin{eqnarray}
\sum\nolimits_i \frac{\partial F_i \rho_\mathrm{ss}}{\partial x_i} - \Omega^{-1} \sum\nolimits_{ij} \frac{\partial^2 D_{ij} \rho_\mathrm{ss}}{\partial x_i \partial x_j}  =  0 ,
\end{eqnarray}
we can obtain a Hamilton-Jacobi equation in the leading order\cite{Graham1984}
\begin{eqnarray}
\mathbf{F} \cdot \nabla\phi  +  \nabla\phi \cdot \mathbf{D} \cdot \nabla\phi  =  0 .
\end{eqnarray}
The Fokker-Planck equation can be thought of as an operation encoding the information of the deterministic driving force $\mathbf{F}$ into the probability distribution $\rho_\mathrm{ss}$.
But, some distortions is also encoded into the latter, due to the fluctuation by diffusion matrix $\mathbf{D}$.
The WKB limit can be regarded as a procedure of reducing the effect of the fluctuation to restore the original intrinsic feature of the deterministic dynamics.
In this sense, the leading order $\phi$ can be thought of as an intrinsic potential.
In additin, the Hamilton-Jacobi equation endows the intrinsic potential with the monotonic decreasing nature of a Lyapunov function:
\begin{eqnarray}
\mathbf{F} \cdot \nabla\phi  =  - \nabla\phi \cdot \mathbf{D} \cdot \nabla\phi  \leq  0 .
\end{eqnarray}

The steady state probability density $\rho_\mathrm{ss}$ in the Fokker-Planck equation and the intrinsic potential $\phi$ in the Hamilton-Jacobi equation are linked by the relationship
\begin{eqnarray}
\rho_\mathrm{ss} = \exp(-\Omega\phi)_{\Omega\rightarrow\infty} .
\end{eqnarray}
In equilibrium systems, the potential function is often known given the interactions. However, the intrinsic potential $\phi$ is usually not known \emph{a priori} for a nonequilibrium system.
Importantly, there exists an intrinsic steady state flux velocity\cite{Zhang2012}
\begin{eqnarray}
\mathbf{V} = (\mathbf{J}_\mathrm{ss}/\rho_\mathrm{ss})_{\Omega\rightarrow\infty}
\end{eqnarray}
perpendicular to the gradient of intrinsic potential\cite{Zhang2012}, i.e., $\mathbf{V} \cdot \nabla \phi = 0$.
In addition, the intrinsic flux velocity is divergent free $\nabla\cdot\mathbf{V} = 0$ and therefore rotational.

It turns out that the nonequilibrium dynamics in the thermodynamic limit is determined by the two orthogonal forces from the gradient of potential landscape characterized by the intrinsic potential $\phi$ and the intrinsic steady state flux velocity \cite{Wang2008,Zhang2012,Wang2015ADP}, i.e.,
\begin{eqnarray}\label{decomposition}
\mathbf{F} = -\mathbf{D} \cdot \nabla\phi + \mathbf{V} .
\end{eqnarray}
The potential landscape through its negative gradient often gives arise to a convergent attraction. For a potential dominated system $\dot{\mathbf{x}} = -\mathbf{D} \cdot \nabla\phi$, the convergent attraction can be against the divergent chaotic behavior.

One can see that the solution $\phi$ should contain an integral constant from the form of Hamilton-Jacobi equation.
Therefore, the effective solution can be written in the form $\phi = \phi(\mathbf{x}) - \phi_\mathrm{min}$ with respect to a ground state potential $\phi_\mathrm{min}$.
Importantly, the system can only be in the ground states $\phi_\mathrm{min}$ in the thermodynamical limit, since the probability density $\rho_\mathrm{ss} = \exp(-\Omega\phi)$ decays exponentially as $\Omega\rightarrow\infty$.
In the ground states, the potential becomes flat locally $\nabla\phi=0$ and hence loses its own function for driving the dynamics.
The system is then governed by the intrinsic flux $\mathbf{V}$ at the ground state.
The chaos may emerge in the flux-dominating regime where $\dot{\mathbf{x}} = \mathbf{V}$ since no convergent gradient attraction is present anymore on the ground states.

The onset of chaos can also be reflected by the underlying potential(Fig.\ref{Fig01}(d)-(f)) and flux(Fig.\ref{Fig01}(g)-(i)).
Here, the intrinsic potential is outlined by an isopotential surface $\phi = \phi'$, and the flux is normalized(details see Appendix B).
%
%
%
%
The system state is attracted by the negative gradient of potential into the ground states surrounded by the isopotential surface $\phi = \phi'$.
The system undergos a phase transition from mono-stable state to chaos state.
The system has a single ground state as shown in Fig.\ref{Fig01}(d) with the lower chemical potential difference.
The system state is attracted into the basin of the intrinsic potential landscape by the negative gradient of intrinsic potential. The intrinsic flux as in Fig.\ref{Fig01}(g) drives the system in a spiraling way approaching to the bottom of the basin.
The increased chemical potential difference drives the system to towards a critical point of phase transition. The system possesses a flat basin of attraction as shown in Fig.\ref{Fig01}(e) and a large region in which the flux becomes important as shown in Fig.\ref{Fig01}(h).
With the further increase of the chemical potential difference, the ground state is expanded into a widely connected region as shown in Fig.\ref{Fig01}(f).
In the chaos stage, the system dynamics is dominated by the intrinsic flux leading to an interlaced and overlapped trajectory as shown in Fig.\ref{Fig01}(c).
Note especially the direction of the flux separation near the vertical axle wire as shown in Fig.\ref{Fig01}(i).
Accordingly, nearby points are diverged repeatedly along the flux lines and eventually become widely separated.

\section{Measure for chaos}

\subsection{Largest Lyapunov exponent}

\begin{figure*}[ht]
\centering\includegraphics[width=4in]{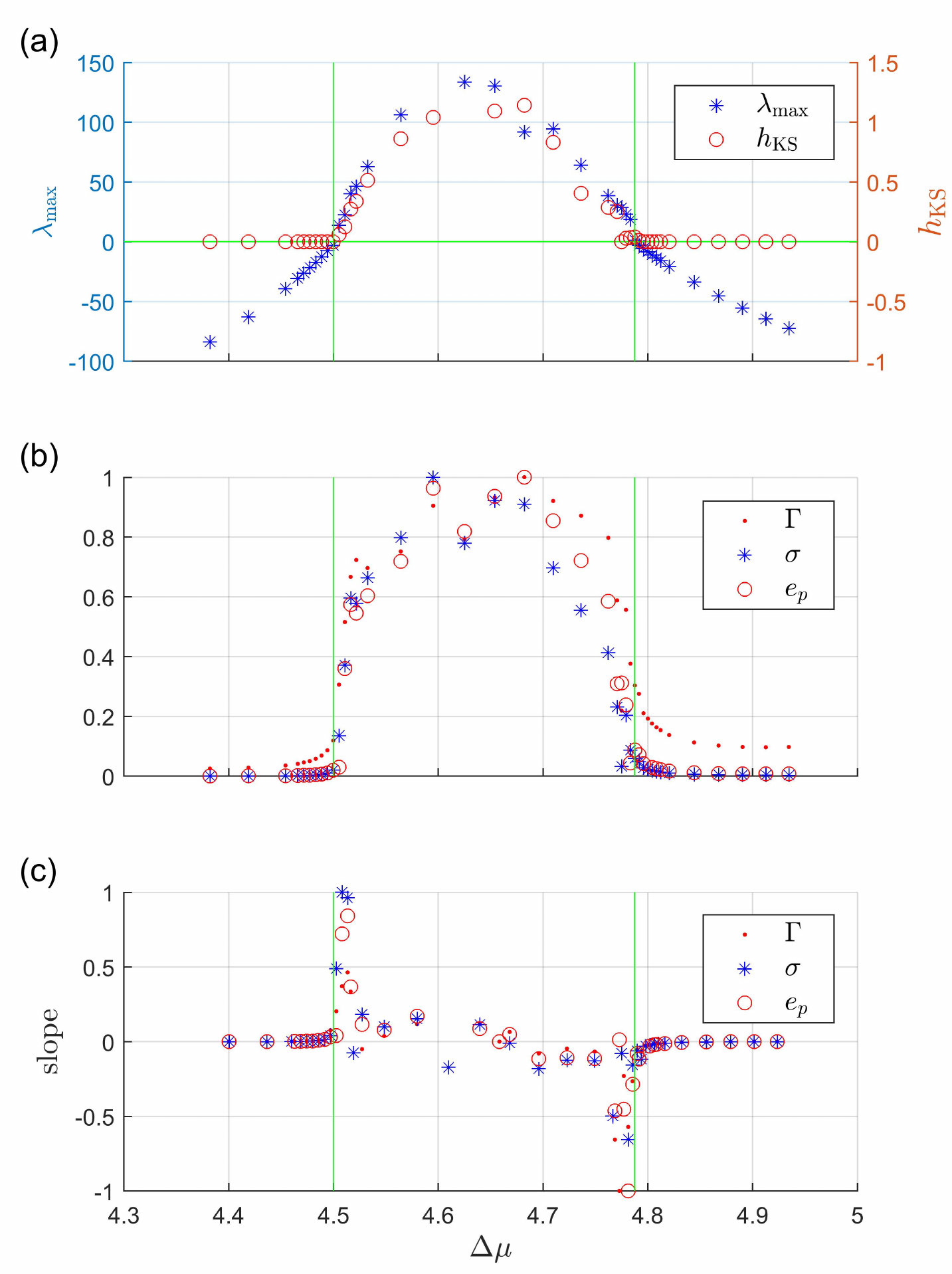}
\caption{
Measures of chaotic degree and its origin.
The largest Lyapunov exponent $\lambda_\mathrm{max}$ measures the crucial feature of chaos, i.e., the sensitivity to initial conditions. The Kolmogorov-Sinai entropy $h_\mathrm{KS}$ quantifying the information creation in the chaotic evolution.
The onset of chaos arises with the increase of chemical potential difference, but the chaos vanishes against the further increase of chemical potential, as in \textbf{a}.
The degree of chaos is correlated strongly with the nonequilibriumness quantified by the average steady state probability flux which is shown in \textbf{b} by dots.
As a direct observation, the time asymmetry of cross correlation as in \textbf{b} marked by asterisks reflects the nonequilibriumness.
The entropy production rate in \textbf{b} shown by circles measures thermodynamical cost for driving the chaos.
It is easy to see all quantities shown in \textbf{b} share the same shape. Significantly, their changes become sharp near the critical points of onset and offset of chaos, as reflected by the slopes in \textbf{c}. Note that there are two distinct picks corresponding to the onset and offset respectively.
}\label{Fig02}
\end{figure*}

The largest Lyapunov exponent $\lambda_\mathrm{max}$ provides a measure for the sensitivity to initial conditions as a crucial kinematic feature of the chaos(details see Appendix D).
It can be calculated by numerical algorithm\cite{Wolf1985}.
The positive Lyapunov exponent implies that the system enters into the stage of chaos.
We have found that the largest Lyapunov exponent increases with the chemical potential difference at the beginning as shown in Fig.\ref{Fig02}(a) denoted by circles, and the positive Lyapunov exponent emerges.
This is in agreement with our common experiences, such as the convection enhanced by the increase of imposed temperature difference.
The chaos appears to emerge upon the increase of chemical potential difference.
%
%
However, the chaos vanishes and the system returns to the mono-stable state with the further increase of chemical potential as shown in Fig.\ref{Fig02}(a).

\subsection{Kolmogorov-Sinai entropy}

Two nearby orbits may appear initially to be the same within certain given accuracy. However, with the chaotic evolution, the orbits separate far enough so that they can be distinguished. In this sense, information is created by the chaos from the perspective of the information theory.
Such a creation of information is measured by Kolmogorov-Sinai entropy $h_\mathrm{KS}$(details see Appendix E).
However, it is difficult to determine its value accurately. Some numerical algorithms were developed to evaluate the Kolmogorov-Sinai entropy.
Here, we use the sample entropy from the time series\cite{Richman2000}. The results are shown in Fig.\ref{Fig02}(a) by the asterisk symbols, where it is normalized by its maximum.
It is easy to see the same trend with the largest Lyapunov exponent represented by the circle symbols.


\section{Physical origin of chaos}

As discussed, the nonequilibriumness can be quantified by the degree of detailed balance breaking in the state space.
%
%
The degree of the detailed balance breaking can be quantified by the intrinsic flux.
Thus, we quantify the degree of the system away from the equilibrium, or the nonequilibriumness, through the average magnitude of the intrinsic flux
\begin{eqnarray}
\Gamma
=
\int |\mathbf{V}| \rho_\mathrm{ss}(\mathbf{x}) d\mathbf{x} .
\end{eqnarray}
The strong nonequilibriumness is associated with the large intrinsic flux which can promote the separation of trajectories.
From the thermodynamical perspective, the dissipation comes from the net input/outpou from the chemical environment reservoirs. 
The entropy production rate(details see Appendix F) is given as
\begin{eqnarray}
e_p
=
\int \mathbf{V} \cdot \mathbf{D}^{-1} \cdot \mathbf{V} \rho_\mathrm{ss}(\mathbf{x}) d\mathbf{x}
\geq 0 .
\end{eqnarray}
related closely to the flux, giving rise to the thermodynamic force in driving the nonequilibrium dynamics.
%
%
It provides a measure of the thermodynamical cost for nonequilibrium systems.
Besides the dynamical features uncovered from the models, the nonequilibriumness can also be inferred by the time reversal asymmetry measured by the difference in cross correlations between two concentration species $i$ and $j$ forward and backward in time
\begin{eqnarray}
\sigma  =  \left| \frac{d [ \langle x_i(0) x_j(\tau)\rangle - \langle x_j(0) x_i(\tau)\rangle ]}{d\tau} \right|
\end{eqnarray}
This provides an direct measure of nonequilibriumness from the experimental observational trajectories \cite{Qian2004,Li2011BJ,Zhang2018JPCB}.

In the thermodynamical limit, the system can only stay in the ground states $\phi = \phi_\mathrm{min}$ with constant probability density $\rho_\mathrm{ss} \sim \exp(-\Omega\phi_\mathrm{min})$, and in which the system dynamics is dominated by the intrinsic flux, i.e.,
\begin{eqnarray}
\mathbf{F} = \mathbf{V} .
\end{eqnarray}
Therefore, the quantification of the nonequilibriumness can be simplified into an approachable form by the trajectory averages
\begin{eqnarray}
\Gamma
=
\lim_{T\rightarrow\infty} \frac{1}{T} \int_0^T |\mathbf{F}| dt .
\end{eqnarray}
where we take the long time average to approach the ensemble average.
%
%
%
In the thermodynamical limit, the entropy production rate can also be converted into the trajectory averages
\begin{eqnarray}
e_p
=
\lim_{T\rightarrow\infty} \frac{1}{T} \int_0^T \mathbf{F} \cdot \mathbf{D}^{-1} \cdot \mathbf{F} dt .
\end{eqnarray}

The nonequilibriumness measured by the average flux $\Gamma$ for dynamics and the entropy production rate $e_p$ for thermodynamics are shown in Fig.\ref{Fig02}(b).
The time reversal asymmetry in cross correlations $\sigma$ between two concentration species $x_1$ and $x_2$ is also shown in Fig.\ref{Fig02}(b).
Note that they are normalized by the respective maximums.
All the above three quantities in Fig.\ref{Fig02}(b) share the similar shape and trend.
With the increase of chemical potential difference, we can see these physical quantities increase when the chemical potential difference is small, while decrease when the chemical potential difference is large.
In addition, we can observe a synchronism between these physical quantities and kinematic chaotic degree measures as shown in Fig.\ref{Fig02}(a).
The flux, the entropy production and the time asymmetry reflect different aspects of the nonequilibriumness of the dissipative system from dynamics, thermodynamics and time direction.
%
%
%
%
%
Importantly, there are distinct sharp changes of the nonequilibriumness physical measures in terms of flux, entropy production and time asymmetry at the critical points of the onset and offset of chaos. Such changes are reflected by the corresponding slopes shown in Fig.\ref{Fig02}(c). Note that the slopes are also normalized by respective maximums. These results demonstrate the nonequilibriumness as the dynamic and thermodynamic origin of the dissipative chaos.
The results also provide alternative physical quantitative signatures for the onset and offset of chaos as a nonequilibrium phase transition.

\section{Discussion and conclusion}

An equilibrium system preserves the detailed balance and its dynamics is determined by the potential alone without the flux.
The flux drives the system into nonequilibrium by breaking the detailed balance. Unlike the gradient force which always attracts the system leading to convergent trajectories, the flux due to its rotational nature tends to drive the system away from the point attractor in a spiral way leading to the divergent trajectories from each other\cite{Wang2015ADP}.
We have noticed that the flux is significant at chaotic phase in a previous work\cite{Li2012JCP}.
In this study, we uncover the nonequilibriumness as the universal origin of the chaos.
From the intrinsic potential-flux decomposition(Eq.\ref{decomposition}), it is easy to see that the chaos is driven by the flux without significant gradient of the potential due to the nearly flat potential landscape near chaotic regime.
The various routes to chaos are in fact different pathways entering into the flux-dominated region.
The sharp changes of the nonequilibriumness of the system, such as the time reversal asymmetry, can provide alternative tracers for the onset and offset of dissipative chaos, which can also become systemic indicators to predict or control the chaos in the engineering applications.

There appears a confusing issue: why the chaos vanishes against the further increase of chemical potential difference?
The chemical potential difference acts as a voltage pump driving the current flow from the higher chemical potential reservoirs to lower one. For a constant resistance, the current increases with the voltage. However, the resistance can change upon the changes of the voltage. The current can decrease if the resistance increases upon the voltage increase. For example, an insulator will have diminishing current upon increase of voltage due to the large resistance. This can help to explain the reason why chaos can vanish against further increase of chemical potential difference. This is because the chaotic system we are studying can change its effective resistance when the chemical potential increases further. This leads to the decrease of the effective current or flux and therefore the associated entropy production rate. In other words, the effective capacitance(resistance) of the system increases(decreases) as chaos emerges and motions are set free. On the other hand, the effective capacitance(resistance) decreases(increases) as chaos disappears and the motions are constrained.

Above all, we can see that the dissipative chaos through the example of chemical Lorentz system is a flux-driven nonequilibrium phenomenon.
The chaos appears with the emergence of the flux-dominated region in the state space. The nonequilibrium flux provides the dynamical origin of dissipative chaos.
The entropy production quantifying the dissipation related to the flux provides the thermodynamical origin of the dissipative chaos.
%
%
The sharp changes of the flux and entropy production rate provide the quantitative indicators for the onset and offset of dissipative chaos.

In the original Lorentz model, only temporal dynamics is focused on for the study of the dissipative chaos. 
By performing the Fourier transformation in space, Lorentz separated the spatial modes and temporal dynamics. 
The spatial dependence in principle can be obtained by truncating the Fourier series to the lowest order of the spatial modes in terms of trigonometry functions. 
In this study, we focused on the temporal behavior of dissipative chaos. The spatial turbulent behavior of dissipative chaos will be explored in the future study.

Our study provides not only an understanding on the origin of dissipative chaos, but also an insight on the nonequilibrium phase transitions.
For example, it can help on the understanding of turbulence as a dissipative chaotic system driven by the nonequilibriumness\cite{Schuster}.
The chaos has also raised concern from ecology for population growth from the competition between reproduction and starvation\cite{May1976}.
In the ecological model, the starvation is often caused by the environmental capacity, such as food. The food intake can be used to measure the degree away from the equilibrium.
In addition, chaos has becomes significant in biological and physiological studies, for instance in the activities of the heart\cite{Poon1997} and the brains\cite{Rabinovich1998,Barak2013}.
The anomalous seemingly chaotic behavior such as epilepsy\cite{Sarbadhikari2001,Iasemidis1996} and heart beat disorder can be associated with the nonequilibrium phase transition.
Our study can also provide novel insight on economics. In particular, after encountering the financial crisis in recent years, the economists called upon the alternative treatment of nonequilibrium chaos beyond the conventional economic theory\cite{Ormerod}.
In addition to these, our study can also inspire studies in other physical and biological fields.

\begin{acknowledgments}
This work was supported by National Natural Science Foundation of China Grants 21721003, Ministry of Science and Technology of China Grants 2016YFA0203200.
\end{acknowledgments}

\appendix

\section{On the model with chaos}

In a specific chemical reaction model, the backward reactions can be taken as very small rates, so that the chemical dynamics is dominated by the forward reactions.
In this case, the specific value of the product concentrations and the backward reaction rates do not appear in the species dynamical equations.
The parameters in the chemical reaction model are taken as follows
\begin{eqnarray}
r_2    &=&  1 ,     \\ \nonumber
r_3    &=&  1 ,     \\ \nonumber
k_1    &=&  0.001 ,     \\ \nonumber
k_2    &=&  1 ,   \\ \nonumber
k_3    &=&  10000 ,   \\ \nonumber
k_4    &=&  0.0001 ,   \\ \nonumber
k_5    &=&  1 ,   \\ \nonumber
k_6    &=&  0.05 ,   \\ \nonumber
k_7    &=&  0.005 ,   \\ \nonumber
k_8    &=&  9900 ,   \\ \nonumber
k_9    &=&  1 ,   \\ \nonumber
k_{10} &=&  0.0133 .
\end{eqnarray}
We take the concentration of the first reactant reservoir $r_1$ as a control parameter.
The corresponding deterministic species dynamics reads from the law of mass action from Eq.\ref{F}
\begin{eqnarray}
\dot{x}_1 = F_1 = &\ &0.001 r_1 x_1 x_2	- 0.1 x_1^2,                          \\ \nonumber
\dot{x}_2 = F_2 = &\ &x_1x_2 - 0.0001x_1x_2x_3 + x_2x_3   \\ \nonumber	
                  &\ &- 0.01x_2^2	- 9900x_2	,		\\ \nonumber
\dot{x}_3 = F_3 = &\ &10000x_3 + 0.0001x_1x_2x_3 - x_2x_3   \\ \nonumber
                  &\ &- x_1x_3 - 0.0266x_3^2 .
\end{eqnarray}
The diffusion matrix reads from Eq.\ref{D}
\begin{eqnarray}
D_{11} = &\ & 0.0005 r_1 x_1 x_2 + 0.1 x_1^2,                          \\ \nonumber
D_{12} = &\ & 0,                          \\ \nonumber
D_{13} = &\ & 0,                          \\ \nonumber
D_{21} = &\ & 0,                          \\ \nonumber
D_{22} = &\ & 0.5 x_1 x_2 + 0.00004 x_1x_2x_3 + 0.5x_2x_3   \\ \nonumber
         &\ & + 0.01x_2^2 + 4950x_2,                          \\ \nonumber
D_{23} = &\ & - 0.00005x_1x_2x_3 - 0.5x_2x_3,                          \\ \nonumber
D_{31} = &\ & 0,                          \\ \nonumber
D_{32} = &\ & - 0.00005x_1x_2x_3 - 0.5x_2x_3,                          \\ \nonumber
D_{33} = &\ & 5000x_3 + 0.00005x_1x_2x_3 + 0.5x_2x_3   \\ \nonumber
         &\ & + 0.5x_1x_3 + 0.0266x_3^2.
\end{eqnarray}
Note that all the terms in the diagonal elements $D_{ii}$ are non-negative as the self-correlations and the diffusion matrix is positive definite.
In addition, the chemical potential difference in this case is given as
\begin{eqnarray}
\Delta\mu = \ln r_1
\end{eqnarray}
where $r_1$ is the concentration of the first reactant reservoirs.

\section{Potential-flux landscape}

The intrinsic potential and flux is obtained by solving the Hamilton-Jacobi equation.
But, the Hamilton-Jacobi equation as a nonlinear partial differential equation is hard to solve.
In an approximation, we replace its solution by the numerical solution of Fokker-Planck equation.

One can see the general Fokker-Planck equation possesses a anisotropic inhomogeneous diffusion matrix
\begin{eqnarray}
D_{ij}(\mathbf{x}) = \sum\nolimits_n v_{ni}v_{nj} w_n(\mathbf{x}) / 2 .
\end{eqnarray}
It leads to also another difficulty in numerical treatment in practice.
So, we use an isotropic homogeneous diffusion to replace the original state-dependent matrix.
It is necessary to notice the isosurface of the potential in Fig.2b is corresponding to the solution of the Fokker-Planck equation
\begin{eqnarray}
\sum\nolimits_i \frac{\partial F_i \rho_\mathrm{ss}}{\partial x_i} - \alpha \sum\nolimits_{ij} \frac{\partial^2 \rho_\mathrm{ss}}{\partial x_i \partial x_j}  =  0 .
\end{eqnarray}
In the numerical calculation, the diffusion parameter is taken as $\alpha = 2\times10^7$.
In the original Fokker-Planck equation
\begin{eqnarray}
\sum\nolimits_i \frac{\partial F_i \rho_\mathrm{ss}}{\partial x_i} - \Omega^{-1} \sum\nolimits_{ij} \frac{\partial^2 D_{ij} \rho_\mathrm{ss}}{\partial x_i \partial x_j}  =  0 ,
\end{eqnarray}
the diffusion matrix is approximately
\begin{eqnarray}
\mathbf{D} =
\begin{bmatrix} 1 & 0 & 0 \\ 0 & 16 & -8 \\ 0 & -8 & 17 \end{bmatrix} \times 10^7.
\end{eqnarray}
So, the parameter $\alpha = 2\times10^7$ is corresponding to an approximate volume $\Omega^{-1}\sim0.1$.

The intrinsic potential $\phi$ contains an integral constant from the form of Hamilton-Jacobi equation, and thus the effective part is the difference
\begin{eqnarray}
\phi = \phi(\mathbf{x}) - \phi_\mathrm{min} .
\end{eqnarray}
In addition, the intrinsic potential can be interpreted by the probability distribution $\phi = -\alpha\ln\rho_\mathrm{ss}$.
So, the integral constant is actually
\begin{eqnarray}
\phi_\mathrm{min} = -\alpha\ln\rho_\mathrm{ss}^\mathrm{max}
\end{eqnarray}
where $\rho_\mathrm{ss}^\mathrm{max}$ is the maximum of the probability density in the state space.
The isosurface for the effective intrinsic potential in Fig.2b is taken as
\begin{eqnarray}
\phi_\mathrm{iso} = 2\alpha .
\end{eqnarray}
Note that the $\alpha$ can be thought of as a noise strength exiting the system.
The system lies in the ground state $\phi_\mathrm{min} = -\alpha\ln\rho_\mathrm{ss}^\mathrm{max}$ without fluctuation.
The isosurface $\phi_\mathrm{iso}$ outlines the states that can be exited by the noise with strength $2\alpha$.
Obviously, the ground state $\phi_\mathrm{min}$ is surrounded by the isosurface $\phi_\mathrm{iso}$.
Furthermore, the intrinsic flux in Fig.2c is also approximated by the solution of the Fokker-Planck equation
\begin{eqnarray}
\mathbf{V} = \mathbf{F} - \alpha\nabla\ln\rho_\mathrm{ss} .
\end{eqnarray}

It is necessary to notice that Fig.2 is a heuristic illustration for the distribution of the potential and flux.
The detailed value used in Fig.2 does not affect the further calculation of other quantities.
The latter is performed by other treatment.

\section{Lyapunov exponent}

The chaos is characterized by the sensitive dependence of system behavior on initial conditions. On average, two infinitesimally closed trajectories separate exponentially fast in the form
\begin{eqnarray}
|\delta \mathbf{x}(t)| \approx e^{\lambda t} |\delta \mathbf{x}(0)|
\end{eqnarray}
with the Lyapunov exponent $\lambda$ providing a measure of such dynamical sensitivity.

More technically, we consider two trajectories $\mathbf{x}(t)$ and $\mathbf{x}'(t)$ with separation $\mathbf{r}(t) = \mathbf{x}'(t)-\mathbf{x}(t)$. For a general dynamics $\dot{\mathbf{x}} = \mathbf{F}(\mathbf{x})$, the separation rate reads
\begin{eqnarray}
\dot{\mathbf{r}}(t) = \mathbf{F}(\mathbf{x}'(t))-\mathbf{F}(\mathbf{x}(t)) \simeq \mathbf{J}(\mathbf{x})\mathbf{r}
\end{eqnarray}
where the Jacobian $\mathbf{J} = \partial\mathbf{F}/\partial\mathbf{x}$. The solution can be formally written as
\begin{eqnarray}
\dot{\mathbf{r}}(t_n) = \left(\prod_i^n \exp[\tau\mathbf{J}(\mathbf{x}(t_i))]\right) \mathbf{r}(0)
\end{eqnarray}
where $\tau$ is infinitesimal time increment.
The global Lyapunov exponents as the chaotic measure are defined as the eigenvalues of the cumulated matrix
\begin{eqnarray}
\Lambda = \lim_{n\rightarrow\infty} \frac{1}{2n\tau} \ln \left[ \left(\prod_i^n \exp[\tau\mathbf{J}(\mathbf{x}(t_i))]\right)^\mathrm{T} \left(\prod_i^n \exp[\tau\mathbf{J}(\mathbf{x}(t_i))]\right) \right]
\end{eqnarray}
which describe a global property with independence of the initial condition.
Usually, the largest Lyapunov exponent $\lambda_\mathrm{max}$ is representative for measure of chaos.

\section{Kolmogorov-Sinai entropy}

We divide the $d$-dimensional state space into boxes with size $\epsilon^d$.
A dynamical trajectory $\tilde{\mathbf{x}}(t)$ can be identified by the box sequence $i_0(0),i_1(\tau),\cdots,i_n(n\tau)$ with the sampling interval $\tau$.
Such a trajectory can be recognized with the information
\begin{eqnarray}
K_n = - \sum\nolimits P_{i_0 \cdots i_n} \ln P_{i_0 \cdots i_n} .
\end{eqnarray}
Accordingly, the additional information $K_{n+1}-K_n$ is needed to predict the next boxes the trajectory passed.
The Kolmogorov-Sinai entropy entropy is defined as the average information
\begin{eqnarray}
h_\mathrm{KS} = \lim_{\tau\to0}\lim_{\epsilon\to0}\lim_{N\to\infty} \frac{1}{N\tau} \sum\nolimits_{n=0}^{N-1} \left( K_{n+1} - K_n \right) .
\end{eqnarray}

Some numerical algorithms were developed to evaluate the Kolmogorov-Sinai entropy.
We use the algorithm called sample entropy from the time series\cite{Richman2000}.

\section{Entropy production rate}

The intrinsic flux gives rise to the dynamical irreversibility exhibiting the asymmetry in the probability between forward and backward trajectories.
The irreversibility in the trajectory probabilities can be obtained by means of the path integral\cite{Jarzynski2006}.
A stochastic trajectory occurs with probability
\begin{eqnarray}
\mathcal{P}[\mathbf{x}(t)] \mathcal{D}\mathbf{x} = \mathcal{P}[\mathbf{x}(t)|\mathbf{x}_0] \rho_\mathrm{ss}(\mathbf{x}_0) \mathcal{D}\mathbf{x}
\end{eqnarray}
where trajectory-dependent transition probability density $\mathcal{P}[\mathbf{x}(t)|\mathbf{x}_0]$ can be obtained by the path integral,
\begin{eqnarray}
\mathcal{P}[\mathbf{x}(t)|\mathbf{x}_0] = \prod_{t=0}^\tau \left[\frac{\Omega}{4\pi|\mathbf{D}|dt}\right]^{1/2} \mathrm{exp}\left[ - \Omega \int_0^\tau \mathcal{L} dt \right] .
\end{eqnarray}
Each trajectory contributes a different weight by the Lagrangian
\begin{eqnarray}
\mathcal{L}  \ = \  \frac{1}{4} \ [ \dot{\mathbf{x}} - \mathbf{F}(\mathbf{x}) ] \cdot \mathbf{D}^{-1} \cdot [ \dot{\mathbf{x}} - \mathbf{F}(\mathbf{x}) ] .
\end{eqnarray}
Obviously, the Lagrangian vanishes
\begin{eqnarray}
\mathcal{L}(\mathbf{x},\dot{\mathbf{x}}) = 0
\end{eqnarray}
along a deterministic trajectory $\dot{\mathbf{x}} = \mathbf{F}(\mathbf{x})$.
For the time reversal trajectory $\tilde{\mathbf{x}}(t) = \mathbf{x}(\tau-t)$, we have $\dot{\tilde{\mathbf{x}}} - \mathbf{F}(\tilde{\mathbf{x}}) = - \dot{\mathbf{x}} - \mathbf{F}( \mathbf{x}) = - 2 \mathbf{F}(\mathbf{x})$ and the corresponding Lagrangian
\begin{eqnarray}
\mathcal{L}(\tilde{\mathbf{x}},\dot{\tilde{\mathbf{x}}}) = \mathbf{F}(\mathbf{x}) \cdot \mathbf{D}^{-1} \cdot \mathbf{F}(\mathbf{x}) .
\end{eqnarray}
The dynamical irreversibility is measured by
\begin{eqnarray}
\frac{\mathcal{P}[\mathbf{x}(t)]}{\mathcal{P}[\tilde{\mathbf{x}}(t)]}
&=&
\frac{\mathcal{P}[\mathbf{x}(t)|\mathbf{x}_0] \rho_\mathrm{ss}(\mathbf{x}_0)}{\mathcal{P}[\tilde{\mathbf{x}}(t)|\tilde{\mathbf{x}}_0] \rho_\mathrm{ss}(\tilde{\mathbf{x}}_0) }   \\ \nonumber
&=&
\frac{\rho_\mathrm{ss}(\mathbf{x}_0)}{\rho_\mathrm{ss}(\mathbf{x}_\tau)}
\mathrm{exp} \left[ \Omega \int_0^\tau [ \mathcal{L}(\tilde{\mathbf{x}},\dot{\tilde{\mathbf{x}}}) - \mathcal{L}(\mathbf{x},\dot{\mathbf{x}}) ] dt \right]     \\ \nonumber
&=&
\mathrm{exp} \left[ \Omega [\phi(\mathbf{x}_\tau) - \phi(\mathbf{x}_0) ]
+ \Omega \int_0^\tau \mathcal{L}(\tilde{\mathbf{x}},\dot{\tilde{\mathbf{x}}}) dt \right]                       \\ \nonumber
&=&
\mathrm{exp} \left[ \Omega \int_0^\tau \mathbf{F} \cdot \nabla \phi dt
+ \Omega \int_0^\tau \mathbf{F} \cdot \mathbf{D}^{-1} \cdot \mathbf{F} dt \right]           \\ \nonumber
&=&
\mathrm{exp} \left[ \Omega \int_0^\tau [\mathbf{F} \cdot \mathbf{D}^{-1} \cdot \mathbf{F} - (\nabla \phi_0) \cdot \mathbf{D} \cdot (\nabla \phi_0) ] dt \right]     \\ \nonumber
&=&
\mathrm{exp} \left[ \Omega \int_0^\tau \mathbf{V} \cdot \mathbf{D}^{-1} \cdot \mathbf{V} dt \right] .
\end{eqnarray}
%
%
On the other hand, it is particularly associated with the trajectory-dependent entropy\cite{Ford2012}
\begin{eqnarray}
\frac{\mathcal{P}[\mathbf{x}(t)]}{\mathcal{P}[\tilde{\mathbf{x}}(t)]}
=
\mathrm{exp}\left[ \int_0^\tau \dot{s}[\mathbf{x}(t)] dt \right] ,
\end{eqnarray}
where $\dot{s}[\mathbf{x}(t)]$ is the trajectory-dependent entropy production\cite{Seifert2005} .
Therefore, we obtain the density of entropy production rate
\begin{eqnarray}
e_p
&=&
\Omega^{-1} \int \dot{s}(\mathbf{x}) \rho_\mathrm{ss}(\mathbf{x}) d\mathbf{x}     \\ \nonumber
&=&
\int \mathbf{V} \cdot \mathbf{D}^{-1} \cdot \mathbf{V} \rho_\mathrm{ss}(\mathbf{x}) d\mathbf{x}
\geq 0 .
\end{eqnarray}


\end{document}